\pdfoutput=1
\RequirePackage{fix-cm}
\documentclass[smallextended]{svjour3}       
\smartqed  
\usepackage{graphicx}
\usepackage{xcolor}
\PassOptionsToPackage{hyphens}{url}\usepackage{hyperref}

\begin{document}

\title{Disinformation in the Online Information Ecosystem: Detection, Mitigation and Challenges
}

\titlerunning{Disinformation Mitigation - An Overview}        

\author{Amrita Bhattacharjee         \and
        Kai Shu \and
        Min Gao \and
        Huan Liu 
}


\institute{A. Bhattacharjee \and H. Liu \at
              Department of Computer Science and Engineering \\
              Arizona State University\\
              Tempe, Arizona, USA\\
              \email{\{abhatt43,huanliu\}@asu.edu}        
           \and
           K. Shu \at
           Department of Computer Science \\
            Illinois Institute of Technology \\
            Chicago, Illinois, USA \\
            \email{kshu@iit.edu}
            \and
            M. Gao \at
            School of Big Data and Software Engineering \\
            Chongqing University \\
            Chongqing, China \\
            \email{gaomin@cqu.edu.cn}
}


\maketitle

\begin{abstract}

With the rapid increase in access to internet and the subsequent growth in the population of online social media users, the quality of information posted, disseminated and consumed via these platforms is an issue of growing concern. A large fraction of the common public turn to social media platforms and in general the internet for news and even information regarding highly concerning issues such as COVID-19 symptoms. Given that the online information ecosystem is extremely noisy, fraught with misinformation and disinformation, and often contaminated by malicious agents spreading propaganda, identifying genuine and good quality information from disinformation is a challenging task for humans. In this regard, there is a significant amount of ongoing research in the directions of disinformation detection and mitigation. In this survey, we discuss the online disinformation problem, focusing on the recent `infodemic' in the wake of the coronavirus pandemic. We then proceed to discuss the inherent challenges in disinformation research, and then elaborate on the computational and interdisciplinary approaches towards mitigation of disinformation, after a short overview of the various directions explored in detection efforts. 

\end{abstract}

\section{Introduction}
\label{intro}
\label{sec:1}

Over the last few decades Artificial Intelligence research has progressed by leaps and bounds, and given the parallel advancements in the development of computational hardware components, better, faster and more advanced methods of using AI are being adopted in different fields including medical science, environmental technology, agriculture, etc. Realizing the vast potential for applying AI to solve socio-economic, political and even cultural issues, experts and practitioners from academia and industry have come together to form organizations and launch projects for this very purpose. A few examples include efforts by Google AI\footnote{\url{https://ai.google/social-good/}}, Microsoft\footnote{\url{https://www.microsoft.com/en-us/ai/ai-for-good}}, International Telecommunication Union(ITU) which is an agency of the United nations\footnote{\url{https://aiforgood.itu.int/}} among others. 

Among the several diverse domains in which AI is being used, one gaining increasing popularity is that of information literacy, detection and mitigation of disinformation, especially on social media~\cite{shudisinformation}. According to a report by the International Telecommunication Union, there were 4.13 billion internet users in 2019, which is more than 50\% of the total 7.7 billion global population. According to the same report by the International Telecommunication Union, among social media platforms, Facebook is one most widely used - by 2.2 billion people in 2018. Along with Facebook, other social media platforms like Twitter, Reddit etc., and personal blogs, opinion based websites are a perfect ground for the spread of false information - \textit{disinformation}. Large scale online disinformation has the potential to shift public opinion, polarize people, and can even threaten public health, democracy and international relations~\cite{gangware2019weapons}~\cite{jackson2018issue}. 

False information and fabricated content on social media exist in different forms and go by different names. \textit{Misinformation}~\cite{wu2019misinformation} is usually referred to erroneous or false information that is disseminated unintentionally, and is often considered as an ``honest mistake". On the other hand, \textit{disinformation} is false information, propagated intentionally with some ulterior motive in mind, be it political, economic, or simply to instigate people. Disinformation can usually be verified to be false and may also consist of true facts expressed in the wrong context in order to mislead readers.  \textit{Fake news}~\cite{deffn} is usually seen as a subset of disinformation and refers to false content or propaganda published in such a way that it appears to be genuine news. \textit{Conspiracy theories}~\cite{bale2007political} have been defined as implausible beliefs and stories that speculate that malicious actors have secretly come together to deceive the public and achieve some goal. Another word used in this context, \textit{hoax} means something that appears to be true but is meant to trick or deceive people. \textit{Troll} or \textit{internet troll}~\cite{troll} is a bot or a human on the internet that tries to instigate and polarize people by spreading inflammatory and highly offensive content and comments. 

\begin{figure}[h]
    \centering
    \includegraphics[width=0.75\textwidth]{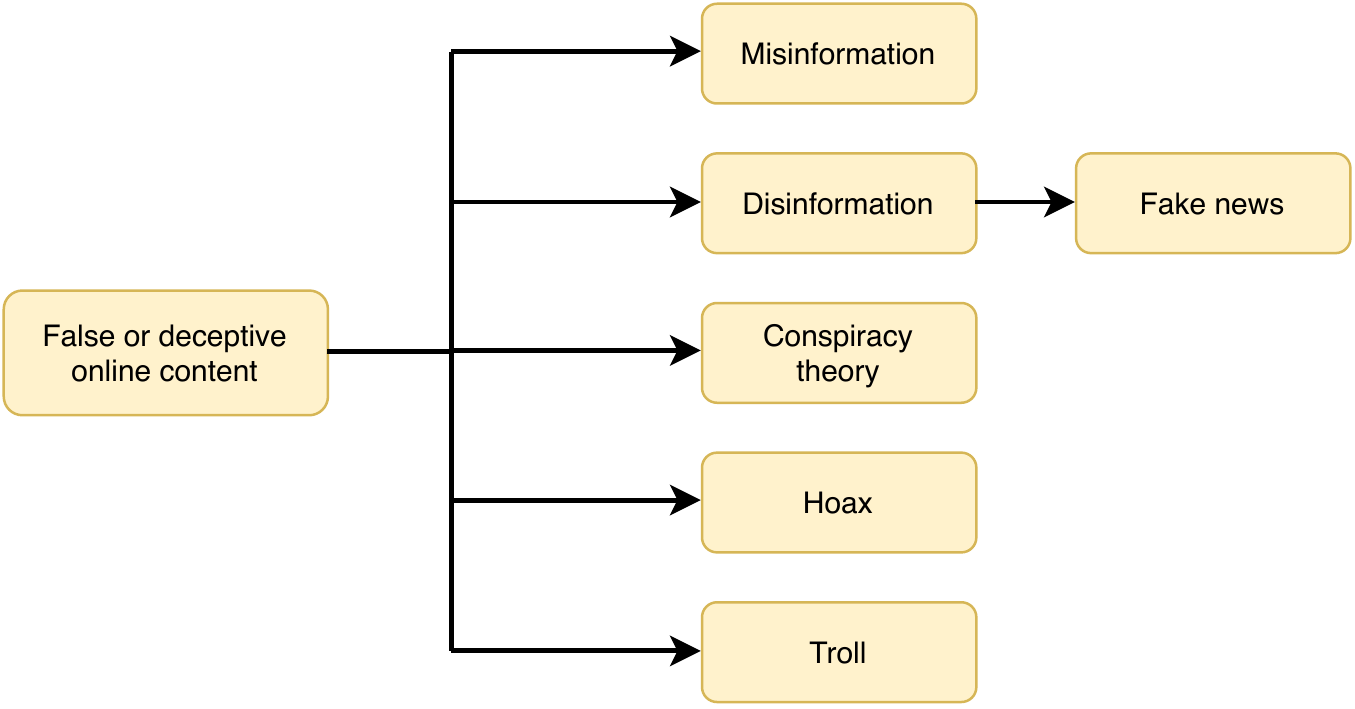}
    \caption{Types of false or deceptive online content}
    \label{fig:my_label}
\end{figure}

In this paper, we focus on the domain of disinformation and discuss past, ongoing and potential future applications of artificial intelligence and interdisciplinary ideas to solve specific problems in detection and mitigation of online disinformation. The rest of the paper is structured as follows: Section \ref{sec:3} presents a case study of disinformation in the context of the coronavirus pandemic. In Section \ref{sec:4} we describe the challenges in disinformation research, particularly in detection and mitigation tasks and discuss some approaches to tackle such challenges. In Section \ref{sec:5} we provide a brief overview on approaches in disinformation detection before discussing computational methods for disinformation mitigation in Section \ref{sec:6}. In Section \ref{sec:7} we elaborate on some interdisciplinary approaches towards disinformation detection and mitigation, before concluding the survey and talking about future work in Section \ref{sec:8}.

\section{Disinformation: A Case Study during the COVID-19 Crisis}
\label{covid}
\label{sec:3}

Although disinformation and fake news have been in existence for the last few decades, the proliferation of social media along with the low cost of generating and spreading content online, have made it increasingly easier for false information to be widely disseminated. Hence, disinformation and fake news now stand as a threat to public opinion and democracy. Recently, the COVID-19 pandemic has brought on a new wave of misinformation and disinformation on social media. The information overload, consisting of both true and false information, on social media has been so overwhelming that the World Health Organization has identified this phenomenon as an `infodemic' ~\cite{infodemic}. According to ~\cite{cons1}, the pandemic has resulted in the spread of 2,000 myths and conspiracy theories regarding the virus. Given the large fraction of people who seek out information on the internet, this is alarming. On 4th May, 2020, a COVID-19 related conspiracy film named `Plandemic' went viral on social media platforms including Facebook and Youtube~\cite{newref1}, spreading dangerous claims that the spread of the virus and the subsequent development of the vaccine is pre-planned by global elites. The video even included apparent frontline doctors claiming that wearing a mask ``activates'' the coronavirus. The spread of such misinformation and disinformation in the context of the coronavirus is especially threatening because it can make people lose trust in scientists, governments, and can gravely undermine a coordinated response to the virus, thereby making it impossible to wipe out or even control the virus. Social media platforms like Twitter and Facebook have been working hard to moderate and remove content that violates their community standards, including coronavirus related misleading posts from politicians~\cite{newref2}. Disinformation related to the coronavirus mostly fall into the following few categories. 

\subsection{Health Related Mis/Disinformation}


The highly infectious nature of the novel coronavirus and the detrimental effects it has on human beings has caused a great amount of panic all across the world, and it is natural for people to be desperate to find cures or promising preventative measures. Although till date there does not exist a cure or vaccine for the virus, individuals and companies have tried to take advantage of the panic and gullibility of consumers and claimed to create/discover products that `cure' or `boost immunity' against the novel coronavirus. Some examples of such false cures include colloidal silver, herbal teas, essential oils, dietary supplements etc. Several groups and companies have been identified and warnings have been issued against these fraudulent activities and claims that have the potential to cause great harm to gullible consumers ~\cite{fakecure}. Dealing with such health-related misinformation or disinformation is quite challenging for a number of reasons. Even official, otherwise credible sources may mistakenly publish wrong misinformation. For example, the US Food and Drug Administration(FDA) had to revoke their authorization of the emergency use of chloroquine and hydroxychloroquine for treating hospitalized COVID-19 patients, as evidence suggested that the risks outweighed the benefits~\cite{fda1}. Furthermore, the President of the United States of America had falsely claimed that ingesting disinfectants including bleach could be effective against COVID-19 ~\cite{bleach3}, and that the drug hydroxychloroquine is a global cure for the coronavirus ~\cite{bleach}~\cite{bleach2}. Both these claims including several other drug related claims making rounds on the internet were fact-checked and debunked. Such dangerous claims have already had detrimental effects and people lacking health-related knowledge have taken harmful steps following such myths, even resulting in death~\cite{bleach4}. 


\subsection{Conspiracy Theories}


There have also been an increasing number of conspiracy theories cropping up regarding the origin and spread of the novel coronavirus. According to a recent study by the Pew Research Center ~\cite{pew1}, 71\% of Americans have been exposed to some conspiracy theory regarding the coronavirus and around a quarter of Americans believe in one of the conspiracies. As shown in the `Plandemic' video, there are conspiracy theories suggesting the virus is man-made and was created by global elites for profit and power. Some conspiracy theories claim that wearing face masks can cause carbon dioxide intoxication, while some view it as an infringement of civil liberties ~\cite{cons2}~\cite{cons3}. According to some conspiracy theorists, 5G radiation helps the spread of the virus, or that the virus is a man-made bioweapon created in a lab. Some others claim that Bill Gates funded the creation of the virus and has intentions of depopulating the world and implanting microchips in people~\cite{cons4}. A large group of people believe that the virus is simply a hoax and is made up or exaggerated by the media, often for political reasons~\cite{cons5}. In the midst of the rapidly evolving coronavirus situation, such conspiracy theories do nothing but fuel the panic and desperation among people who are already desperately seeking answers and explanations. In extreme cases, these conspiracy theories may even damage public trust in the coordinated response to the virus, healthcare regulations, etc. and in the process worsen the situation. 

\subsection{Scams and Frauds}


The coronavirus pandemic has had devastating effects on global economy, and pushed millions of people into unemployment. Governments, including the US government, had arranged for relief/stimulus checks, unemployment benefits, etc. to relieve citizens of some stress during such a difficult time. Unfortunately, scammers have been trying to utilize this situation to their advantage, by tapping into people's despair and desperation and ultimately robbing them in the process. Over the duration of the coronavirus pandemic, malicious groups and individuals have been targeting vulnerable people including older, retired people, leveraging their virus-related fear to push them to make hasty decisions. Many such scams were in the form of scam calls, emails related to stimulus checks, charity scams, and even dubious work-from-home opportunities in the wake of the massive unemployment as a result of the pandemic\footnote{\url{https://www.fcc.gov/covid-scams}}. According to ~\cite{scam1}, 68\% of the over 204,000 complaint calls made to US Federal Trade Commission(FTC) regarding COVID-19 related payments, were to report fraud or identity theft. Fraud websites claimed to offer coronavirus vaccinations, although one does not exist yet, and even offered free coronavirus and antibody tests~\cite{scam2}. Fraudsters even pretended to be contact tracers and contacted gullible people to get their personal information which were then used in health insurance scams~\cite{scam3}. According to ~\cite{scam1}, there have also been instances of scam calls offering help with loan payments and bills, websites selling `miracle' cures for the coronavirus, and even emails with malware links in them. Several government and non-government agencies have put out warnings to make consumers aware of the increased number of scams and frauds during this time, especially COVID-19 related ones.

\section{Challenges in Disinformation Research}
\label{challenge}
\label{sec:4}

Research in the disinformation domain is often challenging and thus, even though there has been a significant amount of work in certain aspects of disinformation, much of the landscape is unexplored and under-explored. There are several issues and nuances that hinder the process, some of which we discuss here:

\begin{itemize}
    \item Broad field - Disinformation and fabricated content have existed in almost all domains, be it political propaganda or fabricated scientific research papers.
    \item Dynamic field - Disinformation is an ever-changing field of research. New problems, challenges and threats keep cropping up every day, with the latest ones being the threat of AI generated news articles and extreme human-like language generation capabilities. So it is important to model future threats and take action accordingly.
    \item Interdisciplinary issue - The problem of fake news and disinformation is heavily an interdisciplinary issue involving computer scientists, political scientists and strategists, policy makers, ethics experts and even psychology and cognitive scientists.
\end{itemize}


Disinformation and false/fabricated content exist in almost all domains and information dissemination platforms, be it social media sites like Facebook and Twitter, or crowd-sourced knowledge bases such as Wikipedia. Since disinformation manifests in such a wide variety of forms, including different modalities such as text, fake videos, fabricated images, claims and images put out of context etc., it is significantly difficult, even as a human being, to understand and comprehend what classifies as disinformation and what does not. Furthermore, pre-existing beliefs, cognitive biases and being confined in echo chambers ~\cite{sunstein2018republic} may also come in the way of making an informed decision regarding the veracity of a claim or post on social media. Hence, as it is quite challenging even for humans to understand and identify disinformation, it is expected that it would be harder for automated detection algorithms to correctly, effectively and efficiently identify disinformation. Furthermore, mitigation is an even harder task because the first step to slowing the spread of a piece of disinformation is detecting it, which itself is very challenging. Apart from that, mitigation also comes with its own set of challenges. We discuss a few of the challenges in \textit{detection} and \textit{mitigation} of disinformation and some approaches to tackle them:

\subsection{Data Challenge}

Disinformation and fake news exist in almost all kinds of domains, be it political, health-related, rumors about celebrities, etc. News from different domains have different styles of writing and even a quite different set of vocabulary. This makes the problem of automatic disinformation detection very challenging, because automatic detection models trained with data from one domain would not perform well when used to detect disinformation from another domain. Furthermore, it is also infeasible to develop separate classifiers for each domain, since it is time and resource consuming, and also because annotated datasets may not be available for all kinds of domains. Annotating datasets is an expensive process and may also require domain knowledge experts. To add to this problem, annotated datasets may have some event or topic specific keywords and phrases, and again, a classifier trained on this dataset would not perform well in other domains where that event or topic is not significant. Furthermore, some news articles may be pure satire without any mal-intention. Disinformation detectors may fail to correctly differentiate a fake article from a satire if it was not trained with that kind of data. Another challenge in the generation of datasets for training, as identified by authors in ~\cite{asr2018data}, is that extracting `true' news articles from reputed news websites and `fake' ones from suspicious websites may not result in a proper dataset for training purposes. This is because even otherwise reputable news sources may post misinformation, and not all news articles on suspicious news websites may be false. Hence, a better way of annotating is required, and instead of relying on the reputation of the news source, the authors have leveraged fact-checking websites to harvest the veracity labels for news articles. 

The problem of domain specific labelled data in disinformation datasets can be partially addressed by using transfer-learning based approaches to train automatic disinformation detectors. Although this idea has not been explored extensively, a few recent works leveraging transfer learning to solve the problem of lack of labelled data for some domain are ~\cite{dhamani2019using}~\cite{guo2020adaptive}. To tackle the issue of domain or event specific features, adversarial techniques could be used to train the disinformation classifier such that it generalizes well to other kinds of domain/event. The objective is to prevent the classifier from learning the domain or event specific features. Following this idea, Wang et al.~\cite{wang2018eann} have proposed an Event Adversarial Neural Network that tries to learn event-independent, shared features instead of event specific ones so that the model performs well on newly emerged disinformation.

\subsection{Early Detection and Effective Mitigation}

Disinformation and deceptive content on social media platforms are designed and created in such a way so as to affect the reader emotionally and trigger a quick response. Sensationalized headlines, `clickbaity' images and language having extreme sentiment are often used to get the attention of the user and get the user to engage with the post. Such false posts may spread like wildfire via readers sharing and re-sharing the post. For example, as we mentioned in Section \ref{covid}, a conspiracy video named `Plandemic' went viral and had 8 million viewers before being taken down~\cite{news2}. The video had numerous false claims regarding the coronavirus pandemic and even claimed that the outbreak was pre-planned by a group of global elites for gaining profit and power. Ever since the video surfaced online, people including influential politicians and athletes shared it and endorsed it. On Facebook the video had around 2.5 million user engagements in the form of likes, comments and shares before it was removed from the platform ~\cite{news3}. Disinformation like these can have disastrous effects such as decreasing public trust in healthcare systems, co-coordinated response to the virus and in turn public health. In high-risk cases like this and otherwise, early detection of disinformation is extremely important in order to minimize the harm caused. However this is an extremely challenging task, in case of both manual and automated detection.
First, manually fact-checking claims for early detection might be challenging for newly emerging incidents as these would require subject matter experts to verify the claims, due to the lack of a pre-existing knowledge base. Furthermore, the process of fact-checking by third-party fact-checkers takes time.
Second, automatic detection would not perform well since the model has not been trained on the newly emerged idea. The performance would be even worse if there are domain inconsistencies between the domain in which the model was trained and the one in which it is being used.
Also, in the early stages of disinformation propagation, there might be a lack of user comments and reactions which may have served as rich sources of auxiliary information for disinformation detection. Early detection of disinformation is also essential for effective mitigation. The sooner a disinformation post can be identified, the quicker a counter cascade can be launched to fact-check the claim and curb the virality of the spread. 

Although this is a relatively under-explored domain, some researchers have proposed frameworks and models to tackle this issue. Liu and Wu ~\cite{liu2018early} proposed a method that detect if a news article is fake based on the characteristics of users on its propagation path. This method requires only the first five minutes of propagation data in order to get a good detection accuracy. To detect fake news articles that have been published on a news website but not disseminated on social media yet, Zhou et al.~\cite{zhou2020fake} proposed a framework that relies on features extracted from the news content - the extracted features have different granularities - syntax-level, lexicon-level, semantic-level etc. To tackle the problem of the lack of user comments as a source of auxiliary information in case of early detection, Qian et al.~\cite{qian2018neural} developed a model that learns to generate user responses conditioned on the article, that were then used in the classification process along with the word and sentence level information from the actual article. An interesting work by Vicario et al.~\cite{vicario2019polarization} identifies topics that might arise in misinformation posts in the future, thus providing some kind of early warning.

\subsection{Challenges in Fact-checking based Approaches}

Researchers and practitioners in journalism domain have been using fact-checking based methods to verify claims, be it written or verbal. 
Manual fact-checking is an extremely slow process and requires domain experts to inspect and investigate the news article and determine its veracity. Given the huge amount of information generated and propagated online everyday, it is impossible for traditional fact-checking to scale up to such a level. Furthermore, in this context early detection becomes a challenge as by the time the veracity of the article has been determined, readers may have already propagated it to several others. Furthermore, it has also been seen that rebuttals and exposure to a fact-checked piece of news after the initial exposure to the false article, does not work as well as we would expect. Ecker ~\cite{ecker2017rebuttals} explains this phenomenon from a psychology point of view, in the context of retraction of faulty claims and how humans react to that. We go into details regarding this topic in Section \ref{sec:int-mit}.
    
Automatic or computational fact-checking ~\cite{ciampaglia2015computational} is a better alternative to manual fact-checking and may potentially alleviate the problem of latency in verifying the truthfulness of claims and articles. However, this approach has its own set of challenges and obstacles. Most automatic fact-checking methods require a pre-existing knowledge base, usually in the form of a knowledge graph, which needs to be regularly updated to ensure the knowledge base is up to date. This is especially difficult in case of newly emerging topics and knowledge, such as the symptoms and severity of COVID-19 in the early stages of the pandemic. Furthermore, natural language processing models trained on a certain corpus of claims may have linguistic biases. Although this can be alleviated by using structured data (subject, predicate, object), getting structured data is difficult. Furthermore, pre-existing knowledge bases may themselves be noisy or biased, based on how these were created. Ensuring the quality of the knowledge-base being used as ground truth is an additional challenge.

\subsection{Multi-modality Approaches}

News articles and posts on social media are usually accompanied by an image or a video related to or supporting the text. Presence of a related image or video alongside the text increases the perceived credibility of the news article or claim, even if the image is a fake or manipulated image. Highly sensationalized posts designed to shift public opinion often leverage carefully manipulated images, videos and even Deepfakes, which are almost indistinguishable to the human eye. Genuine images and videos might even be taken out of context to convey a meaning not originally intended. Hence, this poses an additional challenge in disinformation detection on social media. Given that fabrication of content and hence deceit can occur in so many different forms, it is difficult to design generalizable models to detect disinformation and fabricated content. 

Although individual aspects of the problem, such as fake image detection, have been addressed by many researchers, research along the direction of multi-modal detection is limited and hence has great scope for improvement. Significant efforts in this area include works by Wang et al.  ~\cite{wang2018eann} which uses text and visual feature extractors and then uses the concatenated feature representations to train the fake news detector. However, the correlations between image and text are not captured in this method. In cases where the image and the text in a post are individually both true, but viewing them together may mislead the reader or represent something totally different, it is important to learn the joint representation of text and image. Khattar et al. ~\cite{khattar2019mvae} use a variational autoencoder to learn a shared representation of the text and image, and is able to reconstruct the text and image separately from this shared latent representation, thereby capturing the correlations and dependencies across the two modalities. A similar approach by Zhou et al.~\cite{zhou2020safe} learns the text and image representations separately and then computes similarity between text and visual representations. Identifying a `mismatch' between the modalities helps in detect fake news. Cui et al.~\cite{cui2019same} have incorporated three types of modalities in their model - text, image and the user profile. The proposed fake news detector also incorporates sentiments from user comments to determine if the article/post is fake or not. An interesting approach utilizing knowledge from external real-world knowledge graph, alongside textual and visual features, is that of Zhang et al.~\cite{zhang2019multi}. Video and audio features along with text have been utilized in ~\cite{krishnamurthy2018deep}. Other efforts in multi-modal fake news detection include ~\cite{kang2020multi} ~\cite{singhal2019spotfake}. 

\subsection{Policy Issues and Fairness}

This challenge is mostly unique to the task of disinformation mitigation. After successful detection of a false post on a social media platform, the next step is to decide what corrective action is to be taken. This is especially challenging from the point of view of social media platforms because of debates over the right balance between free speech and protection against false information. Censorship and in particular, social media censorship have been an issue of national and international debate and taking concrete action against online disinformation is difficult. Recently, the US White House declared an executive order~\cite{wh1} against online censorship by social media platforms, on the ground that it violates free speech and the foundations of open debate and even democracy. On the other hand, social media platforms also face immense criticism from a large section of the general public as well as large organization regarding the lousy approach to curb misinformation and disinformation on the platforms~\cite{wh2}. Categorizing a controversial post as disinformation or an opposing viewpoint often requires nuanced debate which makes it difficult for social media platforms to take quick action. Given how rapidly fake news and disinformation can spread on online social media platforms, countries and governments are also concerned about disinformation and propaganda influencing elections. For example, tech giants have been under the radar and the European Union have been putting pressure on these companies to be proactive in regulating disinformation and fake news ~\cite{wh3}~\cite{wh4}. Furthermore, to complicate things, different countries and regions of the world have different cultures when it comes to media, journalism and consumption of news and information.  

In the United States, for example, social media platforms, including their owners and regulators, have often been accused of holding left liberal ideologies and `unfairly' silencing opposing narratives when these had false or deceptive content. Hence, the topic of fairness and political bias comes up and it is an issue of debate. There exists a blurred fine line between free speech, self-expression and potentially causing harm via deceptive or partially false information, especially during important events such as the US Presidential elections and the coronavirus response. Some groups may feel they are being targeted by social media censorship regulations more than others, and hence developing disinformation mitigation policies keeping everyone's interests in mind raises significant fairness concerns.

\section{Detecting Disinformation as a Preliminary Step Towards Mitigation}
\label{sec:5}

In order to mitigate and slow down the spread of disinformation, the first step is to accurately and efficiently identify disinformation. Effective mitigation is only possible if the piece of disinformation is detected early on, i.e. the spread of the post or news article through the network is not yet significant. Given that early and effective detection of disinformation is an indispensable step leading up to its potential mitigation, in this section, we discuss some efforts in different directions of disinformation detection.
Ever since the role of fake news and disinformation in the 2016 US Presidential elections became evident~\cite{allcott2017social}~\cite{shu2019detecting}, there has been an increasing amount of efforts in developing manual and automated ways of detecting disinformation on social media. Several challenges and caveats in the process of effective and efficient detection of disinformation, including the challenges we described in the previous section, have been identified, along with preliminary solutions. In general, the disinformation detection task is to output a `true' or `false' label, given a news article as input. Often there are other auxiliary information along with the news content text. Here we provide a brief summary on some prominent disinformation detection approaches. 

\textbf{News Content and Social Context: }Most earlier disinformation or deception detection approaches used hand-crafted features, often focusing on the textual and linguistic features of the news content. Linguistic features usually included lexical and syntactic features among others~\cite{ott2011finding}~\cite{feng2012syntactic}. Other features were platform-dependent, for example, number of re-tweets and likes on Twitter. Gradually researchers realized that social networks hold a lot of useful information that can be leveraged to improve detection performance. Thus, researchers started using social context ~\cite{shu2019beyond}, user profiles~\cite{shu2018understanding}, user engagement ~\cite{shu2020fakenewsnet}, and relationships among news articles, readers and publishers to detect fake news and disinformation~\cite{della2018automatic}.

\textbf{User Comments: }User comments on social media are rich sources of information when it comes to identifying the truthfulness of an article. For newly emerging topics and events, the ground truth is usually not available. In such cases, user comments can provide weak social signal towards the veracity of the news article, thus helping in early detection ~\cite{shu2020leveraging}. Such a technique can also tackle the challenge of limited labelled data. In the case of early detection, it may so happen that user responses or comments are unavailable since the news article was just posted online. To deal with this situation, authors in ~\cite{qian2018neural} develop a model which learns how users respond to news articles and then generates user responses to aid the detection process. Some recent works in disinformation or deception detection using signals from the crowd are ~\cite{kim2018leveraging}~\cite{tschiatschek2018fake}~\cite{tchakounte2020reliable}.

\textbf{Fact-checking based methods: }Fact-checking ~\cite{vlachos2014fact} based approaches of detecting disinformation may be of two types - \textit{manual} or \textit{automatic}. Manual fact-checking either relies on domain experts to fact-check and verify claims, or crowdsource knowledge from users. Automatic fact-checking based detection approaches use structured knowledge bases such as knowledge graphs to detect the veracity of claims and news articles. This is done by first extracting knowledge (in the form of subject, predicate, object triples) from news articles and then comparing against a pre-existing knowledge graph. Ciampaglia et al. ~\cite{ciampaglia2015computational} used transitive closure in weighted knowledge graphs to derive a value for semantic proximity. A simpler web search based automated fact-checking approach have been proposed by Karadzhov et al.~\cite{karadzhov2017fully}. As explained in ~\cite{graves2018understanding}, automated fact-checking has several challenges - one of the significant ones being the lack of a large-scale supporting knowledge base or knowledge graph.

\textbf{Explainable and Cross-Domain detection: }Some new and interesting areas in fake news detection research are explainable detection and cross-domain detection. With interpretability becoming an important topic of conversation in AI research, explanations behind why a news article was classified as `fake' is an interesting aspect to look into. Efforts in this direction include ~\cite{yang2019xfake}~\cite{shu2019defend}~\cite{reis2019explainable}. Understanding why an article is labelled as fake can help design better detection algorithms. As we mentioned in the previous section, effective cross-domain detection is a potential solution to the labelled data scarcity problem. As shown in ~\cite{janicka2019cross}, naive fake news detectors do not perform well across domains and hence, there is a lot of scope for research in this area.

\section{Computational Methods for Mitigating Disinformation}
\label{sec:6}

Given the unpredictable and noisy online information ecosystem, and the ever increasing ease of creating and disseminating content, mitigating the spread of false content and claims has emerged to be another important task. Some studies have shown that fake news and disinformation spread faster and wider than true news ~\cite{vosoughi2018spread}~\cite{news1}. This is mostly because disinformation generates more extreme reactions in readers, thus making then engage with the content more. Since social media ranking algorithms are designed in such a way so as to promote posts and content that get more user engagement, disinformation posts gain more visibility due to high user activity. Algorithms that govern the visibility of articles and posts on social media, including recommendation and ranking based algorithms need to be designed with this in mind. Furthermore, effective mitigation of disinformation campaigns on social media also requires early detection of such disinformation. In this section, we discuss a few of the computational approaches in which mitigation of disinformation is addressed.

\subsection{Network Intervention}

A social media platform may be represented as a network where nodes are the users (i.e. accounts on that social media platform) and edges are the connections between them (such as `friendship'). Disinformation on social media spreads in the form of a cascade, where the `infection' is initiated by a few nodes that created or posted the false content, and then it is propagated by other nodes that get influenced by this piece of disinformation. If the disinformation is capable of influencing the exposed users effectively, such a spread can grow exponentially and hence, early detection and mitigation of the disinformation campaign is of paramount importance. One way of slowing down the spread of the disinformation cascade is to launch a counter cascade, for example, one that contains the fact-checked version of that disinformation article. Such a problem is often referred to as the influence limitation or influence minimization problem.
In the influence limitation problem, the objective is to find a (or the smallest) set of nodes in the network where a counter-cascade could be initiated such that the effect of the original cascade could  be minimized. Although finding the optimal set of seed nodes for counter campaign propagation has been proven to be NP-hard, however, approximation algorithms have been developed for some variants of the problem ~\cite{budak2011limiting}~\cite{nguyen2012containment}~\cite{luo2014time}. In case of disinformation mitigation, the objective of this problem is to `immunize' or inoculate as many nodes as quickly as possible, in order to minimize the harm caused by the spread of the piece of disinformation. Budak et al.~\cite{budak2011limiting} have described a greedy approach for a variant of the influence limitation problem, using carefully chosen heuristics, to arrive to an approximate solution to the problem. This proposed algorithm works on a discrete time model as opposed to a continuous time model used in ~\cite{luo2014time}. While most previous works considered only two cascades, ~\cite{tong2018misinformation} tackled the problem of multiple cascades and also introduced the concept of `cascade priority' whereby a node in the network has varying probabilities of getting influenced by each of the cascades. In a recent work ~\cite{hosni2020minimizing}, the authors have considered the problem of cross-platform misinformation propagation and proposed a greedy approach based on survival theory to solve this issue.

\subsection{Content Flagging}

Social media platforms that host user-generated content often give users the option to `flag' or `report' some content, if it is offensive, harmful and/or is false information etc. In platforms where there is significant freedom in generating and disseminating content, a centralized system of moderating such user-generated content is extremely challenging, if not impossible, due to the sheer volume of the content and also the domain-knowledge that might be necessary to debunk false claims. Hence, most social media platforms leverage the readers and the audience to maintain the quality of content online, whereby a user can raise `flag' the content if it violates the community guidelines of that platform. In the specific case of flagging content as `false information', these platforms often use 3rd party fact-checkers to judge the veracity of the claim and take required action. If the claim is proven to be false or harmful, the social media platform may remove the post entirely or reduce its reach. Some platforms also notify users of the fact-check performed and provides context as to why sharing/interacting with the post is not ideal ~\cite{mosseri2016news}~\cite{lyons2017news}.
While some political and non-political organizations have been putting pressure on social media platforms to take proactive steps in controlling the spread of illegal content online ~\cite{techgiant}, several other organizations have reprimanded the same social platforms when disinformation and false claims were flagged and removed from the platform, on the grounds that such false claims and opinions posted by people are not `illegal', and this kind of censorship by social media platforms threatens free speech ~\cite{factchk}. Hence, content flagging and removal for mitigating disinformation is quite challenging and there are a lot of opportunities for developing better solutions.  

\subsection{User Susceptibility}

Another direction explored in mitigation of disinformation is analyzing user behavior and identifying susceptible or gullible users, i.e. users who are most vulnerable to consumption and propagation of fake news or false claims ~\cite{rajabi2019user}~\cite{sameki2017crowd}. Authors in ~\cite{rajabi2019user} attempted to identify groups of vulnerable users on Twitter when it comes to fake news consumption, using an unsupervised approach to cluster users into 3 categories - good, malicious, and vulnerable/naive - based on user, content and network features. Recent studies have shown that certain groups of users, for example, older people~\cite{newref3}, might be more susceptible to fake news and disinformation. Such identification of vulnerable groups may help in performing targeted fact-checking, or using specialized but ethical ways of delivering the true information or facts to these groups, after a previous exposure to fake news. 

\section{An Interdisciplinary Look at Disinformation}
\label{sec:7}


As we mentioned in Section \ref{challenge}, disinformation research requires a significant amount of interdisciplinary collaborations. Methods and models in computer science that deploy artificial intelligence and machine learning to detect disinformation and slow the spread of disinformation are only a small part of the solution. Such techniques need to be augmented by knowledge from other domains. A lot can be learned from domains such as social science, psychology and cognitive science regarding how and why rational human beings fall prey to fake news and disinformation, what psychological processes trigger responses to and engagement with disinformation, etc. Although researchers have started exploring such ideas, there is a lot more left to understand and learn from, and such auxiliary knowledge from other disciplines can help in the design and development of better detection and mitigation solutions.

Furthermore, the problem of disinformation is not just a technical one. Public outreach programs with convincing messages and campaigns need to be organized to address the issue of disinformation, so that the lay person can make better decisions with respect to the quality of news and information they consume.

\subsection{Understanding and Identifying Disinformation Interdisciplinarily}
\label{sec:int-mit}

In order to tackle the problem of online disinformation and deception at the grass roots level, several studies have been performed to understand the cognitive and psychological processes at play when an individual interacts with and believes in a piece of fake news, and in general, what makes humans vulnerable to disinformation. As described in ~\cite{time}, when presented with too much information and resources, the human mind naturally tends to use cognitive shortcuts or `heuristics' in order to make decisions, instead of critically analyzing each and every piece of information. Such is the case in today's online information ecosystem where people are inundated with news, opinions, viewpoints, that are often extremely contradictory, thus making the task of assessing the veracity of such articles increasingly difficult. The idea of cognitive shortcuts and heuristics in the context of credibility evaluation has been mentioned in ~\cite{metzger2013credibility}, and the authors explain how the use of such heuristics may often lead humans to make incorrect judgements and decisions. Among the heuristics described by authors in ~\cite{metzger2013credibility}, some are \textit{reputation heuristic} - users judge the credibility of a new article by how reputed or recognized the news source is, \textit{endorsement heuristic} - users tend to believe a news article to be credible if a known and trusted acquaintance shared and hence endorsed the piece of news, and \textit{self-confirmation heuristic} or confirmation bias - users tend to gravitate towards news and information sources that align with their pre-existing beliefs.

Another interesting idea behind vulnerability to disinformation is that of \textit{Dual Process Theory} ~\cite{pennycook2017perspective}, according to which, the human mind has two ways of thinking and processing information - the first is an automatic (unconscious) process that requires little effort, and the second is an analytic (conscious) process that requires significant amount of effort. Quick, automatic processing, similar to using heuristics for information processing may also mislead humans and make them misinterpret the credibility of news. Motivated reasoning, that is reasoning and rationalizing in such a way so as to align with prior beliefs, is also another reason behind the gullibility of humans to disinformation~\cite{cog1}~\cite{cog2}. Nowadays social media platforms often use 3rd party fact-checkers to verify the truthfulness of news articles and claims being shared on the platforms. If a disinformation post is fact-checked and found to be false, several social media platforms, such as Twitter and Facebook, usually display a tag or a disclaimer, mentioning the result of the fact-check~\cite{mosseri2016news}. However, studies and research have shown that such rebuttals are ineffective~\cite{ecker2017rebuttals} and may even trigger readers of the disinformation piece to strengthen their beliefs in the original false narrative. Some researchers ~\cite{cook2015misinformation}~\cite{nyhan2010corrections} explain how rebuttals and retractions of misinformation can even reinforce people's belief in the misinformation by the `backfire' effect. In a less extreme scenario, such corrective retractions may not have any effect on the reader due to the lag between the exposure to the original piece of misinformation and the retraction, since by that time the readers already have established mental models to make sense of that piece of misinformation~\cite{thorson2016belief}. To complicate the matter even further, experiments have shown that exposure to `disputed' flags or fact-checks on disinformation headlines seems to increase the perceived truthfulness of news headlines that have not been fact-checked and hence do not have a visible flag~\cite{pennycook2020implied}. Hence, disinformation mitigation is an extremely challenging task and requires knowledge and concepts from various disciplines and domains. Knowledge from cognitive science has also been borrowed and used in disinformation detection. An example is ~\cite{kumar2014detecting}, where the authors have tried to detect disinformation by using cues such as consistency and coherency of the message, source credibility and general acceptability.

\subsection{Interdisciplinary Efforts in Disinformation Mitigation}


Apart from the technical approaches towards disinformation mitigation described above, there have also been efforts in other domains, such as, psychology and cognitive sciences. Fake news and disinformation is a global phenomenon and with today's ease of online content generation and propagation, it is becoming more and more important to spread awareness regarding the prevalence of fake news and how to best avoid it. With around 4.57 billion active internet users worldwide(as of July 2020)~\cite{intuser}, there is a growing need to accept that false information and deception is here to stay, and to spread awareness among the public about the existence of fake news and how to spot fake news. Furthermore, several studies and surveys have shown that humans are extremely susceptible to fake news; even individuals who claim to be immune to fake news~\cite{singapore}~\cite{psych}. A 2019 study conducted by the Stanford History Education Group~\cite{stanford} showed that students were unable to assess the credibility of online sources and were susceptible to misinformation. A shocking 90\% of the students failed on 4 out of the 6 tasks they were asked to perform, thus demonstrating the need for media literacy. Many organizations, academics and non-profits have contributed to efforts in this aspect and here we mention some of the notable ones. Some organizations dedicated to promoting media literacy among the public are Media Education Lab\footnote{\url{https://mediaeducationlab.com/}}, Media Literacy Now\footnote{\url{https://medialiteracynow.org/}}, The National Association for Media Literacy Education\footnote{\url{https://namle.net/}} and Center for Media Literacy\footnote{\url{http://www.medialit.org/}}. Several journalistic standards also exist to prevent misleading or false information from being created, and also to maintain ethics. The Trust Project\footnote{\url{https://thetrustproject.org/}} is a consortium of around 120 news organizations and they have developed 8 `Trust Indicators' in an effort to establish trustworthiness, transparency and accountability of news sources. Their objective is to re-establish public trust in quality news sources through these metrics or indicators of trust and accountability. Campaigns to increase public awareness regarding prevalence of fake news and pointers on how to spot fake news are conducted in many countries~\cite{mlcamp1}~\cite{mlcamp6}. Similar campaigns are also being included in the curriculum of schools to educate children from a young age about how to interact with online media~\cite{mlcamp2}~\cite{mlcamp3}~\cite{mlcamp4}~\cite{mlcamp5}. Another media literacy initiative is Drog~\cite{drog}: apart from educational programs and campaigns, they have also designed a game that helps the player build psychological resistance to disinformation by learning how disinformation is created and disseminated ~\cite{roozenbeek2019fake}.

\section{Conclusion and Future Work}
\label{sec:8}

In this paper we gave a brief overview of the current situation of disinformation research, especially disinformation on online social media platforms. We discussed the various challenges in disinformation detection and mentioned some of the ways researchers have tackled such challenges. We then focused on disinformation mitigation efforts, both from a computer science and an interdisciplinary point of view. 

As we mentioned previously, much of the disinformation research landscape is under-explored and there is immense scope for future work. With recent advancements in large, highly expressive language models and automatic text generation capabilities, large-scale disinformation attacks seem to be a potential threat that researchers should prepare for. OpenAI's GPT-2 and GPT-3, that came out very recently, are extremely powerful. Malicious agents can fine-tune such models to automate the generation of false or misleading news articles~\cite{open2}~\cite{open3}. Given how human-like the generation usually is, readers may get easily convinced that the machine-generated article is a legitimate news article. When combined with fabricated images, videos and Deepfakes, this presents an added dimension of difficulty to designing defense methods. Defending against such attacks is a promising new area of research that needs focus at this moment. 

Furthermore, more interdisciplinary research in the mitigation of disinformation is necessary. Although there are independent works on human gullibility and susceptibility to disinformation, these concepts can be utilized in computer science to design better recommendation or ranking algorithms to rank news articles on social media platforms. Current recommendation and ranking algorithms on social media platforms tend to selectively display news stories to users that align with their pre-existing beliefs, thus exacerbating the problem of echo chambers, increasing polarization among the population and ultimately making the problem of disinformation mitigation even more challenging. Hence, there is a need to design better algorithms to ensure the users are exposed to a fine balance between desirable news articles and articles with opposing viewpoints.

%

\bibliographystyle{spmpsci}      
\bibliography{reference.bib}   

\end{document}